\documentclass[reprint,aps,prb,floats]{revtex4-1}

\usepackage[dvips]{color}
\usepackage{graphicx}
\usepackage{bm}
\usepackage{textcomp} 

\begin{document}

\title{\textcolor{blue}{Saturation and bistability of defect-mode intersubband polaritons}} 

\author{Simone Zanotto}
\email{simone.zanotto@polimi.it}
\affiliation{NEST, Istituto Nanoscienze - CNR and Scuola Normale Superiore, P.za S. Silvestro 12, 56127 Pisa, Italy}

\author{Federica Bianco}
\affiliation{NEST, Istituto Nanoscienze - CNR and Scuola Normale Superiore, P.za S. Silvestro 12, 56127 Pisa, Italy}

\author{Giorgio Biasiol} 
\affiliation{Laboratorio TASC, CNR--IOM, Area Science Park, 34149 Trieste,
Italy}

\author{Lucia Sorba}
\affiliation{NEST, Istituto Nanoscienze - CNR and Scuola Normale Superiore, P.za S. Silvestro 12, 56127 Pisa, Italy}

\author{Alessandro Tredicucci}
\affiliation{NEST, Istituto Nanoscienze - CNR, P.za S. Silvestro 12, 56127 Pisa, Italy}
\affiliation{Dipartimento di Fisica ``E. Fermi'', Universit\`a di Pisa, Largo Pontecorvo 3, 56127 Pisa, Italy}

\date{\today}

\begin{abstract}

In this article we report about linear and nonlinear optical properties of intersubband cavity polariton samples, where the resonant photonic mode is a defect state in a metallo-dielectric photonic crystal slab. By tuning a single geometric parameter of the resonator, the cavity Q-factor can reach values as large as 85, with a consequent large cooperativity for the light-matter interaction. We show that a device featuring large cooperativity leads to sharp saturation, or even bistability, of the polariton states. This nonlinear dynamics occurs at the crossover between the weak and the strong coupling regimes, where the \textit{weak critical coupling} concept plays a fundamental role. 

\end{abstract}

\maketitle

\section{Introduction}

Intersubband polaritons are quasiparticles stemming from the strong light-matter interaction between an intersubband plasmon excitation and an electromagnetic resonance. Since the theoretical prediction\cite{LiuPRB1997} and the first experimental observation\cite{DiniPRL2003}, several geometries for the electromagnetic resonator have been implemented, either in order to extend the operating frequency into the terahertz range\cite{TodorovPRL2009, GeiserAPL2010, StrupiechonskiAPL2012} and squeeze the electromagnetic field in strongly subwavelength volumes\cite{TodorovPRL2010, FeuilletPalmaOE2012}, or to engineer the photonic dispersion\cite{ZanottoAPL2010, ManceauAPL2014}. These efforts are motivated by fundamental questions and by applied physics issues. On the one side, lowering the operating frequency and hence entering the terahertz range enabled to achieve wider Rabi splittings, which are connected to a larger emission rate of photon pairs from the nonadiabatically-modulated polariton vacuum, in a phenomenon reminiscent of the \textit{dynamical Casimir effect}\cite{CiutiPRB2005, DeLiberatoPRL2007}. On the other side, mid-infrared or terahertz polaritonic devices, where the resonator possibly features a very small modal volume, are expected to lead to novel polariton lasing mechanisms, overcoming the low radiative efficiency of quantum cascade lasers\cite{DeLiberatoPRL2009, DeLiberatoPRB2013}. 

It is well known that the photonic resonator influences the light-matter coupling by means of two key properties: the modal volume $V$ and the quality factor $Q$. The modal volume enters the definition of the vacuum Rabi frequency via the vacuum electric field amplitude, $\Omega \propto E_{\mathrm{vac}} \propto \sqrt{1/V}$; at a first glance, larger splittings $\Omega$ require smaller modal volumes. However, this applies for a \textit{single} quantum emitter. In the intersubband polariton case,  the matter excitation is a \textit{bright} superposition of $N$ single-particle intersubband excitations, and the Rabi splitting undergoes a superradiant-like enhancement which leads to the final expression $\Omega \propto \sqrt{N/V} \equiv \sqrt{n_{\mathrm{3d}}}$, since $N$ coincides with the number of electrons in the active volume. Better yet, a more refined formula also takes into account the \textit{overlap} $\psi$ between the resonant field distribution and the quantum wells\cite{ZanottoPRB2012}; anyway, the final expression $\Omega \propto \psi \sqrt{n_{\mathrm{3d}}}$ does not depend on $V$. Rather, the push towards smaller modal volumes deals with the concern of avoiding to electrically pump the inactive \textit{dark states}, whose number scales as $V \cdot n_{\mathrm{3d}}$, in an electroluminescent polariton device\cite{DelteilPRB2011}.

Less attention has instead been devoted to study how the cavity quality factor influences intersubband polariton devices; perhaps, this is due to the lack of direct mechanisms for tuning the $Q$-factor of LC resonators and of patch cavities. However, we recently reported about two mechanisms for easily tuning the radiative $Q$-factor of photonic crystal-based resonators, both in the direction of smaller and of larger $Q$'s\cite{ManceauAPL2013, ZanottoJOSAB2014}. It is widely recognized that the quality factor plays a significant role in cavity quantum electrodynamics and in photonics, hence, it is expected to strongly influence intersubband polariton devices. For instance, the well-known photonic concept of \textit{critical coupling} has been recently extended to the strong coupling regime, leading to \textit{strong critical coupling} and perfect polariton absorption\cite{ZanottoNatPhys2014}. The strong critical coupling condition involves the cavity $Q$ by requiring the damping rate \textit{matching} between the intersubband and the photonic resonances, through the usual relation for the photonic damping rate $\gamma_c = \omega_c/2 Q$. 

In this article we will instead explore the effect of the damping rate \textit{mismatch}, relying on the tunability of defect-mode photonic crystal resonators\cite{ZanottoJOSAB2014}. In a low-intensity transmission experiment, defect-mode intersubband polaritons  appear as usual as a doublet of spectral features, whose contrast is governed by the damping rate mismatch. When a high-intensity, monochromatic source is employed, a peculiar sharp saturation and bistable behaviour is instead predicted. These phenomena, which may constitute the physical ground for actual mid-infrared photonic devices, have a neat theoretical interpretation as well: in the final part of this article we will connect the observed phenomenology to the presence of a large \textit{cooperativity} value, which -- more than the damping rate mismatch -- turns out to be the proper theoretical ground for the predicted nonlinearities.

\section{Linear optical properties of defect--mode intersubband polaritons}

A schematic of the samples we analyzed is given in Fig.~1.
\begin{figure}
 \includegraphics{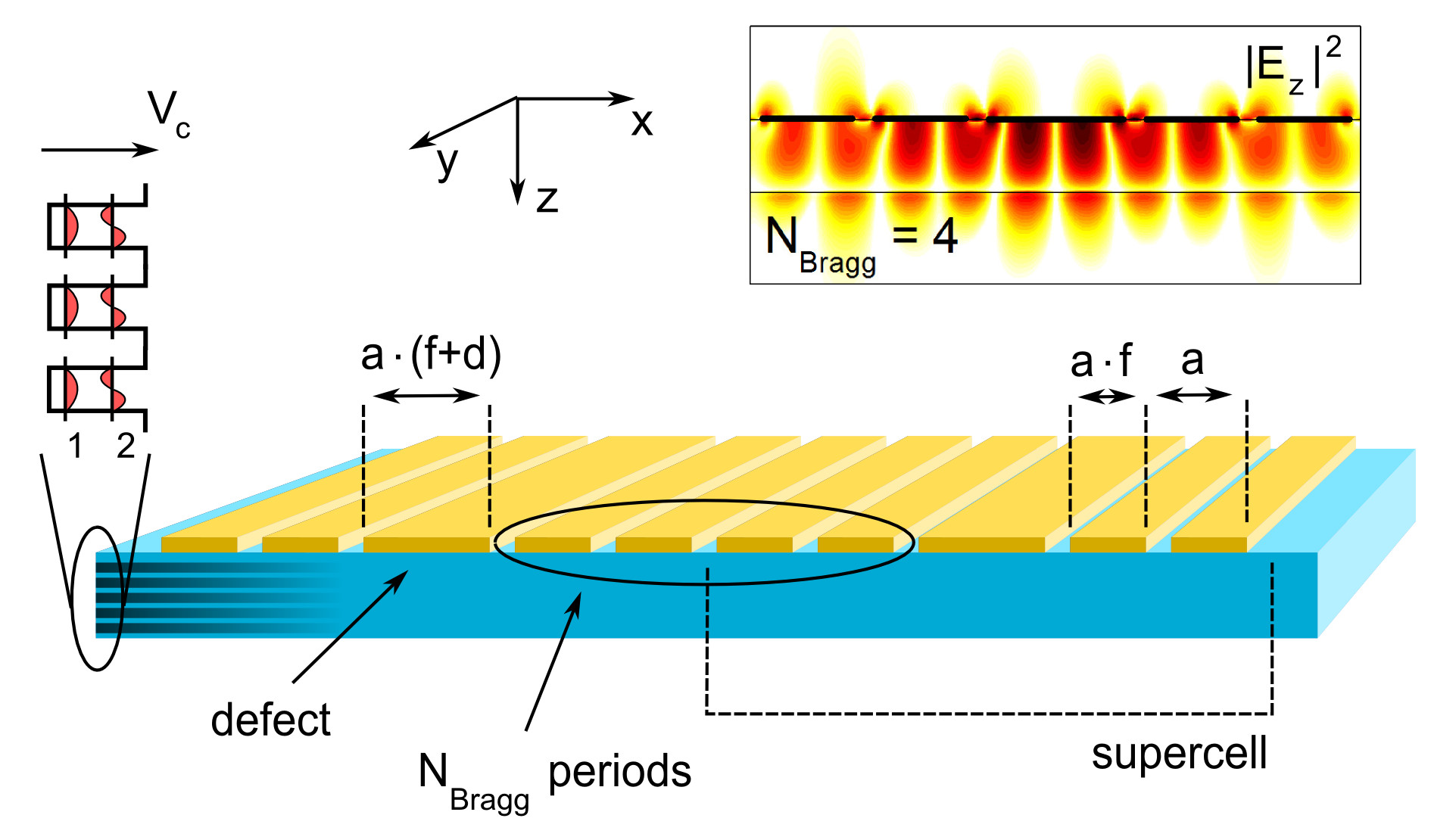}
\caption{Schematics of the photonic crystal sample embedding a multi-quantum well structure. The conduction band profile $V_c$ and the electronic energy sublevels are sketched on the left. When excited at resonance, the defect-like photonic mode has a field distribution with a maximum below the central stripe in the supercell. The plotted field component is $|E_z|^2$; this interacts with the intersubband polarization eventually leading to defect-mode intersubband polaritons.}
\end{figure}
 They consist of a semiconductor membrane patterned with gold stripes, following a supercell scheme where $N_{\mathrm{Bragg}}$ identical stripes (Bragg mirror) are interleaved with a larger stripe (defect). This metallo-dielectric photonic crystal resonator,  whose fabrication procedure is described elsewhere\cite{ZanottoPRB2012R}, features a resonance whose field distribution has a maximum below the defect stripe. As reported in Ref.~\onlinecite{ZanottoJOSAB2014}, the resonance quality factor is governed by the parameters $d$ and $N_{\mathrm{Bragg}}$; on the three samples we realized, we kept fixed $d = 0.4$ and employed $N_{\mathrm{Bragg}} = 2$, $4$, and $6$. As it can be noticed in the figure inset, the resonant field distribution is mainly overlapped with the semiconductor membrane, which is nanostructured implementing a multi-quantum well (MQW, consisting of 65 repetitions of 6.8/15\, nm $\mathrm{GaAs}$/$\mathrm{Al}_{0.33}\mathrm{Ga}_{0.67}\mathrm{As}$ well/barrier pairs). Silicon impurities in the wells provide a $n$-type doping that partially fills the first electronic subband.
The photonic period $a$ of the three samples is calibrated in order to bring the defect mode in resonance with the intersubband transition: we employed $a = 2.96$, $3.06$, and $3.1\ \mu \mathrm{m}$ in correspondence to $N_{\mathrm{Bragg}} = 2$, $4$ and $6$. The fill factor $f = 0.8$ is always employed. 

\begin{figure}
 \includegraphics{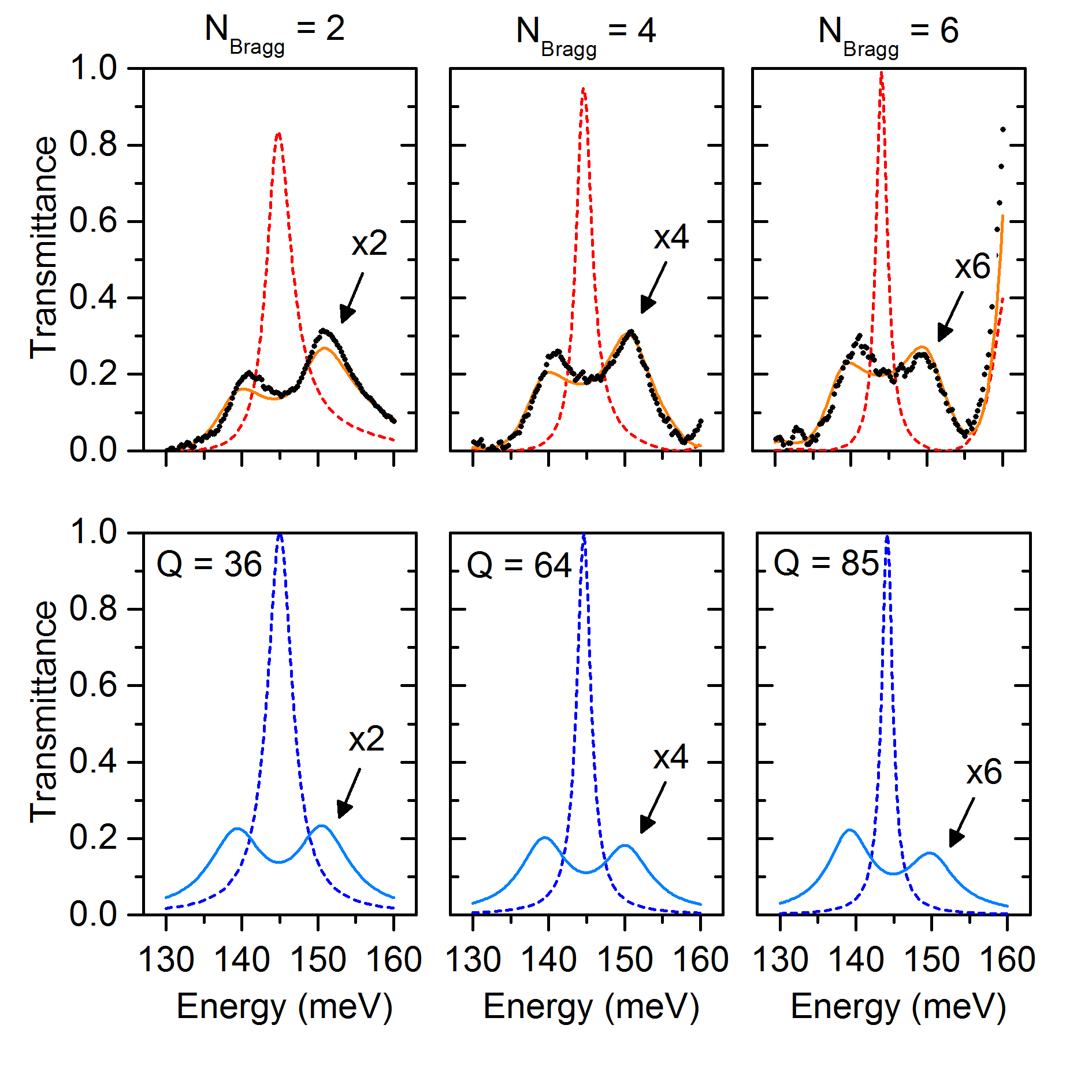}
\caption{Transmittance curves of defect-mode photonic crystals, with active (doped) or inactive (undoped) quantum wells (QWs). Dashed lines: bare photonic resonances (undoped QWs); solid lines, polaritonic resonances observed when the QWs are doped. The lines in the top panels are obtained from a rigorous coupled wave analysis method, while those in the bottom panels follow from a coupled-mode theory model. Samples with different $N_{\mathrm{Bragg}}$ result in different quality factors $Q$ of the photonic mode, and in different contrasts of the polaritonic features. Black dots in the upper panels are experimental data measured by Fourier transform infrared spectroscopy; the experiment only concerned a sample with doped quantum wells.}
\end{figure}

The low-intensity linear optical response is measured by means of Fourier-transform infrared spectroscopy (FTIR), using as source the beam from a glo-bar lamp, polarized along $y$. The transmittance spectra (at normal incidence) are reported as black dots in the upper panels of Fig.~2, where the clear polaritonic doublet is well reproduced by the theoretical traces. The latter are obtained by solving Maxwell's equations with the rigorous coupled-wave analysis method (RCWA), where the input parameters are the geometry and the dielectric functions. The gold is assumed to have $\varepsilon = -4000$, while the MQW is modeled as an effective dispersive anisotropic medium\cite{ZanottoPRB2012R}. In essence, the $z$ component of the dielectric tensor is that of an ensemble of Lorentz oscillators, embedded in a background medium with permittivity $\varepsilon_{bg} = 10.05$, and characterized by a resonance frequency $\hbar \omega_{12} = 144.9\ \mathrm{meV}$, by a damping rate (i.e., half linewidth) $\hbar \gamma_{12} = 6.4\ \mathrm{meV}$, and by an oscillator strength which is proportional to the surface charge difference between the two subbands in the QWs: $\Delta n = n_1 - n_2 = 2.39 \cdot 10^{11}\ \mathrm{cm}^{-2}$. Actually, since the excitation intensity is small and the number of thermally activated electrons is negligible even at room temperature, one has $\Delta n = n_1 = n^{(0)}$, where $n^{(0)}$ is the doping-induced charge density. The values reported above follow from a fitting procedure performed on the polaritonic spectra of the upper panels of Fig.~2, and are in good agreement with both the nominal growth parameters and the observations from independent measurements (i.e., multipass absorption). Our choice of a Lorentz-like broadening for the intersubband transition, rather than a Gaussian or a Voigt one, leads to a satisfactory fit of the experimental data for the coupled system within the measurement errors, and it can be shown that it does not invalidate the main conclusions reported in the remainder of the article. In the upper panels of Fig.~2 we report as red dashed lines the result of a further RCWA simulation, which was performed under the condition $\Delta n = 0$. From the width of the observed transmission peak, which corresponds to the bare photonic crystal cavity resonance, the cavity $Q$ is obtained: the values are reported in the lower panels of Fig.~2, and correspond to the damping rates $\hbar \gamma_c = 2$, 1.13, and 0.85 meV, respectively.

As $Q$ increases, the contrast of the polaritonic features is strongly reduced, reaching values as low as 5\,\% for $N_{\mathrm{Bragg}} = 6$. This is attributed to the increasing mismatch between the damping rates of the intersubband transition and of the photonic cavity: as recently demonstrated by means of a coupled-mode theory (CMT) model\cite{ZanottoNatPhys2014}, the difference $\gamma_{12} - \gamma_c$ plays a key role in the physics of strongly coupled dissipative light-matter systems. When $\gamma_{12} = \gamma_c$ and $\Omega > \gamma$, the \textit{strong critical coupling} condition is fulfilled and the \textit{absorbance} of the polaritonic system is the largest; meanwhile, the contrast of the transmittance doublet is maximised. Here, instead, when $N_{\mathrm{Bragg}}$ is increased, the mismatch $\gamma_{12} - \gamma_c$ progressively grows, the energy feeding into polariton states becomes less and less efficient, and the transmittance contrast is reduced. The ability of the CMT picture to quantitatively capture the transmittance contrast decrease can be gained by observing the lower panels of Fig.~2. Now, the input parameters of the theory are no longer the sample geometry and the material consitutive relations; rather, the model relies on the few physically meaningful parameters $\gamma_c$, $\gamma_{12}$ and $\Omega$. It should be noticed that, while increasing $N_{\mathrm{Bragg}}$ strongly affects the damping rate mismatch and hence the transmittance contrast, the polaritonic splitting essentially does not feel the effect of $N_{\mathrm{Bragg}}$. This is a clear proof that only the field overlap factor $\psi$, and not the resonant field enhancement -- the latter being connected to the $Q$-factor -- influences the intersubband polariton splitting. Indeed, the good agreement between the experiment and the models relies on the relation
\begin{equation}
\Omega = \psi \sqrt{\frac{\pi e^2 \Delta n }{ \varepsilon_w m^* L_{\mathrm{per}}}}
\end{equation}
where $\varepsilon_w$ is the well material permittivity, $m^*$ is the effective conduction band electron mass, $L_{\mathrm{per}}$ is the MQW period, and $\Delta n$  is the surface charge difference between the first and second subband. 

\section{Non--linear optical properties of defect--mode intersubband polaritons}

The independence of the Rabi splitting upon the $Q$-factor is a consequence of the general features outlined at the beginning of the article, which are valid in the linear response regime. When exciting the sample with a sufficiently intense light beam, non-linear phenomena like polariton saturation (\textit{polariton bleaching}) are known to occur\cite{ZanottoPRB2012R}, and it may be guessed that the saturation threshold depends on the cavity $Q$. We performed a bleaching experiment on the defect-mode polariton samples, looking for the intensity-dependent collapse of the transmission doublet, when intense mid-infrared pulses (generated by a system consisting of an optical parametric amplifier and a difference frequency generator) is employed instead of the glo-bar lamp source. As reported in Fig.~3 (a), by tuning the incident intensity with a set of neutral density filters the polariton splitting progressively vanishes as long as a sufficiently large pulse intensity is employed. 
\begin{figure}
 \includegraphics{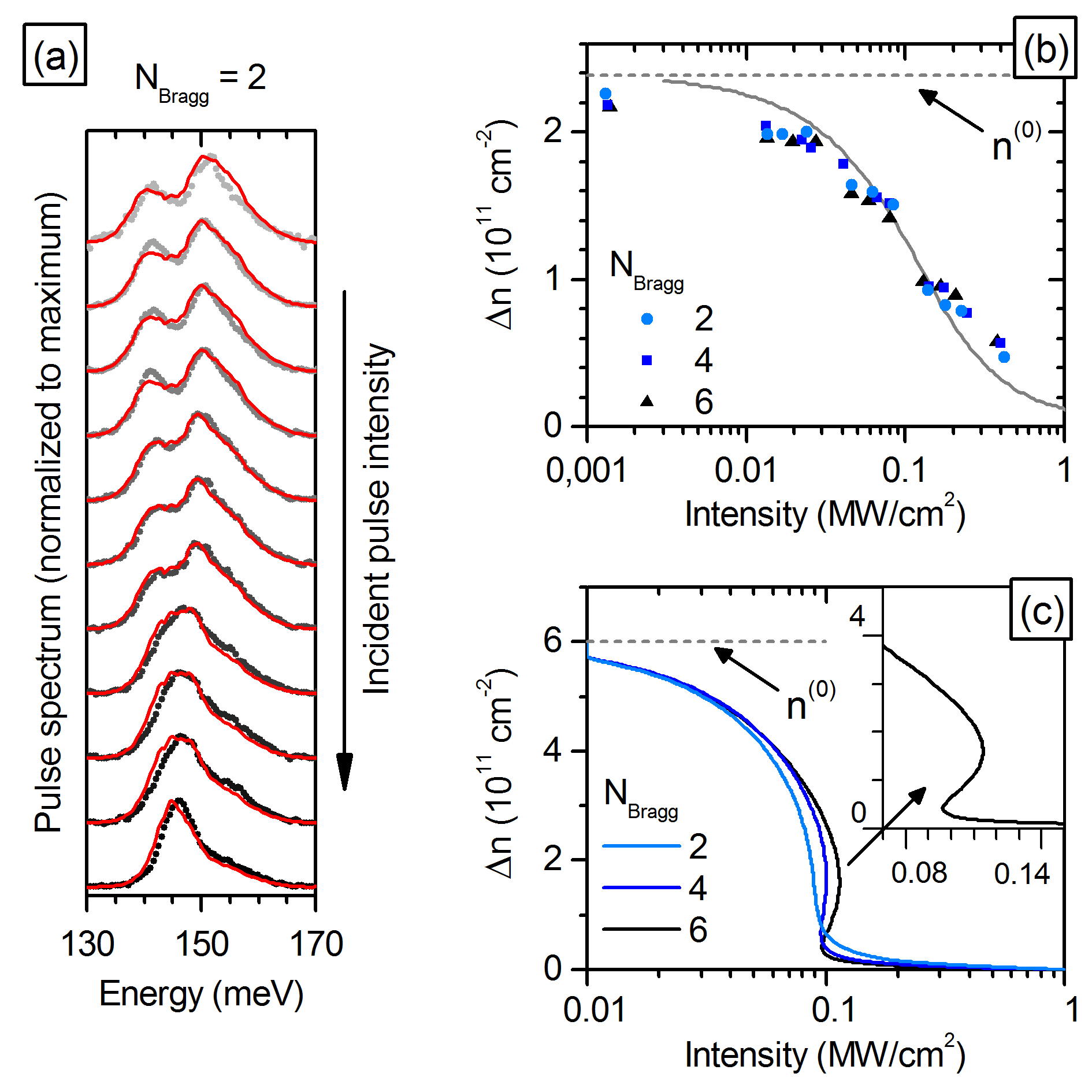}
\caption{(a) Spectra of a broadband ultrafast mid-infrared pulse ($E \in [130, 170]\ \mathrm{meV}$) transmitted through the defect-mode photonic crystal sample with $N_{\mathrm{Bragg}} = 2$. Similar traces are obtained for the $N_{\mathrm{Bragg}} = 4, 6$ samples. The dotted traces correspond to experimental data, while the red traces are obtained by fitting through rigorous coupled-wave analysis; a constant offset has been applied for clarity. (b) Dots: surface charge difference extracted from the spectra as in (a), and theoretical trend obtained by the rate-equation model (line). No appreciable dependence upon the number of Bragg periods is observed. (c) Saturation dynamics predicted by the rate-equation model for a narrowband excitation (i.e., $E \simeq 145\ \mathrm{meV}$. A sudden threshold is now observed, evolving in a bistability loop for large $N_{\mathrm{Bragg}}$'s.}
\end{figure}

In order to quantitatively compare the saturation thresholds occurring on the different samples, the transmitted pulse spectra have been fitted by RCWA, where the surface charge difference $\Delta n$ was regarded as a fitting parameter.
These values are reported as dots in Fig.~3 (b), where it is observed that, contrary to what might have been expected, the polariton bleaching occurs at the same intensity for all samples. This phenomenon can be interpreted by means of a rate-equation model. The idea is that each absorbed photon promotes an intersubband excitation, which eventually decays through a non-radiative mechanism\footnote{The radiative decay path is neglected since the non-radiative decay rate of intersubband excitations is much larger than the radiative one.}. In formulas,
\begin{equation}
 \frac{n_2}{\tau_{12}}\frac{\hbar \omega}{I} = A(\Delta n,\omega)
\end{equation}
where $n_2$ is the charge density in the second subband, $\tau_{12} \simeq 1/\gamma_{12}$ is the decay time, $I$ is the incident intensity and $A$ the sample absorbance, which is dependent on the photon frequency $\omega$ and on the surface charge difference $\Delta n$. In the RCWA modelling, the dependence $A(\Delta n)$ occurs via the effective dielectric function\cite{ZanottoPRB2012R}. Finally, the constraints $\Delta n = n_1 - n_2$ and $n_1 + n_2 = n^{(0)}$, where $n^{(0)}$ is the static charge density provided by the doping, allows to solve Eq.~2 self-consistently and to obtain the relation $\Delta n (I)$. The above model applies straightforwardly to the case of a monochromatic excitation; when dealing with a broadband light pulse like that employed in the bleaching experiment\footnote{Mid-infrared pulses with duration of $\simeq 100\ \mathrm{fs}$, and spectral width $\simeq 30\ \mathrm{meV}$, have been employed.}, one could instead replace the fixed-frequency absorbance with an averaged quantity $\bar{A}(\Delta n)$, where the spectral averaging window is given by the light pulse spectral width. By employing this procedure it turns out that the dependence $\Delta n (I)$ is essentially the same for the three samples, hence uncorrelated with the cavity $Q-$factor. This theoretical result is reported in Fig.~3 (b) as a solid line, which accurately reproduces the experimental trend\footnote{Owing to technical difficulties, in the experiment we only had access to the relative values of the incident power, hence the measured points are not linked to a calibrated intensity scale. However, a horizontal rigid shift of the experimental points is sufficient to overlap with the theoretical curve.}.

The reason why samples with different cavity $Q$-factors essentially behave in the same way even in the non-linear response regime lies in the fact that the excitation employed here is broadband, and the integration over the whole absorption spectrum implies a kind of ``sum rule''. Indeed, the function $\bar{A}(\Delta n)$ is essentially independent on the value of $N_{\mathrm{Bragg}}$, and hence on the cavity $Q$. In addition, $\bar{A}(\Delta n)$ is a monotonic function, resulting in a smooth transition between the unbleached and bleached states. We now wonder which is instead the system's response to a narrowband excitation. If the energy corresponding to the bare photonic resonance is chosen in Eq.~2, the saturation curves reported in Fig.~3 (c) are obtained. Now, the smooth, monotonic saturation curve observed above is replaced by a steep function, possibly showing a hysteresis loop. While bistable behaviours in the exciton-polariton framework have already been reported in the literature\cite{TredicucciPRA1996, BaasPRA2004, BajoniPRL2008}, this is the first prediction of intersubband polariton bistability.
These findings rely on the absorption spectra calculated with the RCWA, and hence are based on actual parameters of photonic crystal membrane devices which can be realized in practice. Such devices differ from the ones we fabricated only for the doping, which needs to be increased to $n^{(0)} = 6 \cdot 10^{11}\ \mathrm{cm}^{-2}$.
It is worth noticing that the sudden decrease in $\Delta n$ is followed by a sharp increase in the transmittance, which switches from values close to zero up to near unity. This behaviour could be at the base of an efficient, ultra-thin mid-infrared saturable absorber, with potential applications in connection to mode-locking of the quantum cascade laser\cite{PaiellaScience2000, MenyukPRL2009, AnisuzzumanOE2010}. 

While the predictions based on RCWA are an invaluable tool for designing an actual device, an analysis based on the coupled-mode theory unveils the basic physical mechanism lying behind the operation of the sharp saturable absorber, or bistable device. Better yet, it allows to extend the operation principle to different frameworks, other than intersubband polaritons, where the coupled-mode model can be applied. For what concerns our two-port photonic crystal slab system, the key is the absorption formula for the coupled light-matter system\cite{ZanottoNatPhys2014}
\begin{equation}
A(\Delta n, \omega) = \frac{2 \gamma_c \gamma_{12} \Omega^2}{\Delta \omega^4 + \Delta \omega^2 (\gamma_c^2+\gamma_{12}^2-2\Omega^2)+(\Omega^2+\gamma_c \gamma_{12})^2}
\end{equation}
where $\Omega$ is the light-matter coupling constant, connected to $\Delta n$ through Eq.~1; $\Delta \omega = \omega - \omega_0$ is the detuning from the cavity and material resonances, assumed to be coincident. This analytical expression, plugged into Eq.~2, justifies both the smooth and the sudden saturation behaviours observed in the above. Indeed, if the expression in Eq.~3 is averaged with respect to $\Delta \omega$, a monotonic function of $\Delta n$ is recovered, confirming the ``smoothing'' picture introduced in the case of a broadband pulse. In addition, it turns out that this smoothed function has a weak dependence upon the damping rates, hence confirming that in the broadband excitation regime the cavity quality factor does not play a significant role. 

\begin{figure}
 \includegraphics{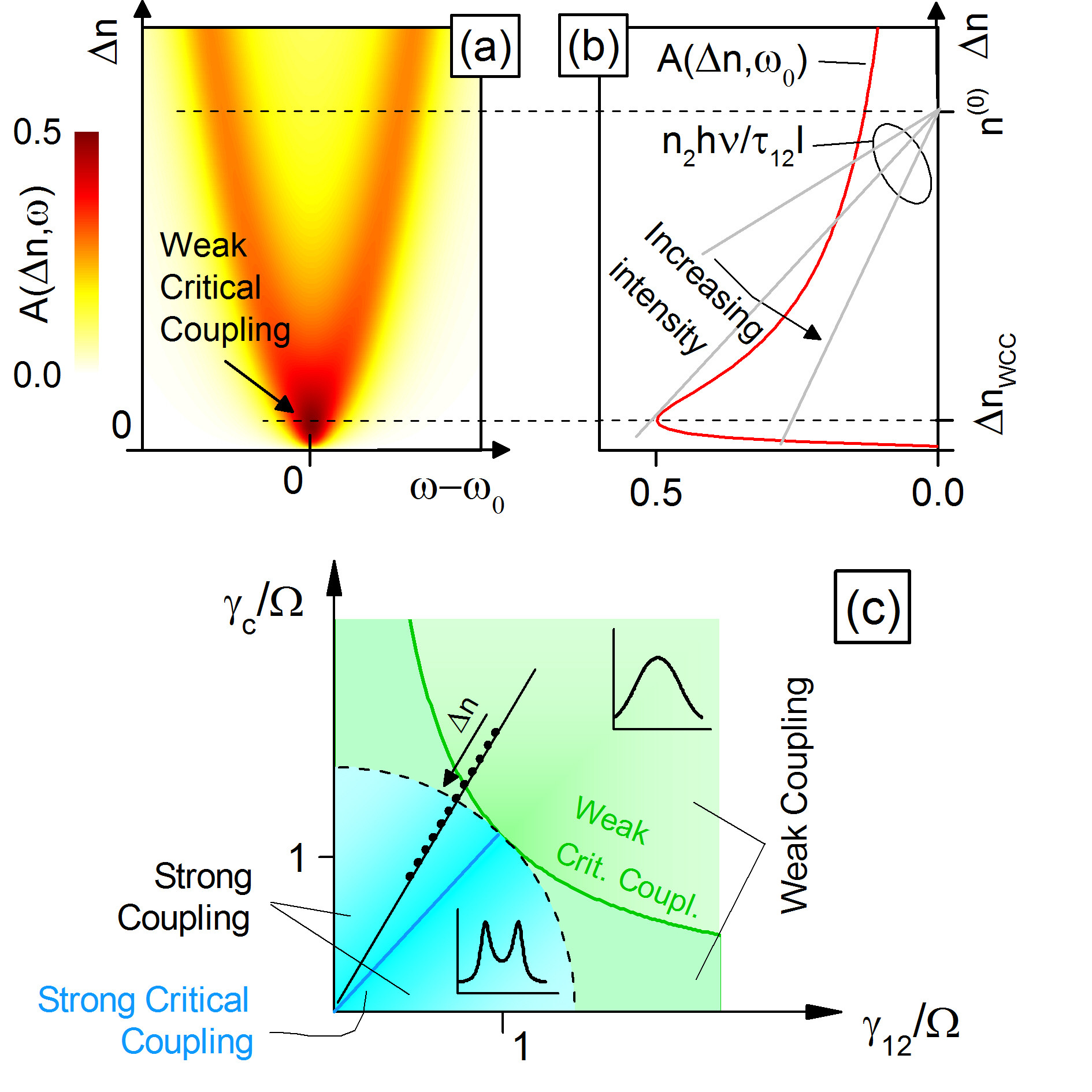}
\caption{Graphical interpretation of saturation and bistability mechanism, from coupled-mode theory and rate equation models. (a) Absorbance spectrum of the photonic cavity as a function of the quantum well charge difference $\Delta n$. It always exist a $\Delta n$ such that $A(\omega_0) = 1/2$; this is the \textit{weak critical coupling condition} usually expressed in terms of Rabi splitting and damping rates as $\Omega^2 = \gamma_c \gamma_{12}$ (where $\Omega \propto \sqrt {\Delta n}$). (b) Graphical solution of Eq.~2, at $\omega = \omega_0$. Saturation and bistability occur in the region of weak critical coupling. (c) Saturation and bistability as a crossover between weak and strong coupling, represented on the coupled mode theory phase diagram. For fixed cavity and transition damping rates, the incident intensity drives the system's working point along the dotted line, across the weak critical coupling. }
\end{figure}

When instead a narrowband excitation tuned at $\omega_0$ is considered, the function $A(\Delta n, \omega = \omega_0) $ has to be employed. This function is no more monotonic with respect to $\Omega$ (and to $\Delta n$), hence triggering the sharp saturation and bistability that was predicted in Fig.~3 (c) through the RCWA-calculated absorption spectra. These features can be immediately grasped by a graphical analysis of the problem, which is proposed in Fig.~4. Panel (a) represents the absorbance calculated from coupled-mode theory, given in Eq.~3. The absorbance spectrum, as a function of $\omega$, shows either one or two peaks, and it always exist a $\Delta n$ such that $A = 1/2$. This surface charge difference is that for which the system is in \textit{weak critical coupling} (WCC), usually expressed in terms of the cavity and transition damping rates, and of the light-matter coupling constant, as $\Omega^2 = \gamma_c \gamma_{12}$. It is actually in vicinity of the weak critical coupling that the relevant nonlinear dynamics, i.e., the saturation and bistability, takes place. This follows straightforwardly from the graphical analysis of Eq.~2, proposed in Fig.~4 (b). As the intensity is increased, the straight line (i.e., the right hand side of Eq.~2, expressed in terms of $\Delta n$ and of the static surface charge $n^{(0)}$) sweeps through the weak critical coupling absorption maximum, possibly with multiple crossings. The presence of multiple crossings is ruled by the relative position of the surface charge difference which gives WCC ($\Delta n_{\mathrm{WCC}}$) and the position of $n^{(0)}$. A completion of the graphical picture is given in panel (c), where we sketch on the coupled-mode theory phase diagram the path explored by the system throughout the bleaching process. Since the cavity and transition damping rates are assumed not to change, the working point always lies on a straight line passing from the origin. The position of the working point is determined by $\Omega$, hence by $\Delta n$, and may lie either in the strong coupling region (where the absorption spectrum is double-peaked) or in the weak coupling one (where the spectrum is single-peaked); close to this boundary (circular dashed line) the weak critical coupling locus is found (green hyperbola). We also notice that the \textit{strong critical coupling} locus (blue segment), i.e., the locus where the absorption spectrum has two peaks at the maximum value of $1/2$, does not appear to play a significant role in connection to the polariton saturation. 

It is instead the concept of cooperativity\cite{Kimble}, $C = \Omega^2/2 \gamma_c \gamma_{12}$, which rules polariton saturation and bistability. Indeed, the weak critical coupling -- and not the strong critical coupling -- is directly connected to the cooperativity: at WCC, $C_{\mathrm{WCC}} = 1/2$. As already hinted by the graphical analysis of Eq.~2, given in Fig.~4 (b), the presence of bistability is governed by the relative position of $n^{(0)}$ and $\Delta n_{\mathrm{WCC}}$, connected to the ratio $C_0/C_{\mathrm{WCC}} = 2 C_0$, where  $C_0 = \Omega_0^2/2 \gamma_c \gamma_{12} \propto n^{(0)} / \gamma_c \gamma_{12} $ is the cooperativity of the unbleached sample. In essence, in order to observe sharp polariton saturation or even bistability, the unbleached sample must exhibit a sufficiently large cooperativity\footnote{The connection between cooperativity and bistability has ben already reported in the literature; see, e.g., A. Desaix \textit{et al.}, Eur.~Phys.~Journal B, \textbf{6}, 183 (1998); J. Sauer \textit{et al.}, Phys.~Rev.~A \textbf{69}, 051804 (2004). The role of cooperativity in connection with intersubband polaritons has instead been highlighted in Y. Todorov \textit{et al.}, Phys.~Rev.~B \textbf{86}, 125314, although within a linear response framework.}. Since in the intersubband polariton framework the transition damping rate cannot be tailored to wide extents, the ability to tune the cavity damping rate (i.e., the cavity $Q$--factor) turns out to be the cornerstone towards the observation of intersubband polariton bistability in samples with a reasonable value of the ground-state charge density $n^{(0)}$.

In conclusion, we realized an intersubband polaritonic device where the cavity mode is a defect-type resonance in a metallo-dielectric photonic crystal slab. With this class of resonators, it is possible to achieve large quality factors, and hence small cavity decay rates, which are strongly mismatched with respect to the intersubband transition decay rate. While this mismatch limits the visibility of polaritonic features in a linear spectroscopic experiment, it enables certain nonlinearities which occur at large intensity, like sharp saturation or bistability of intersubband polaritons. Besides deserving an interest as mid-infrared optical components, these phenomena have a neat interpretation in terms of fundamental physics, and can potentially be exported to a multitude of systems. Indeed, they follow from simple assumptions -- a rate-equation model in conjunction with a coupled-oscillator model -- and only involve the basic physical concepts of cooperativity and of weak critical coupling. 

The authors gratefully acknowledge Raffaele Colombelli and Riccardo Degl'Innocenti for stimulating discussions, and Ji-Hua Xu for the precious support with the ultrafast laser source. This work was partially supported by the European Research Council through the advanced grant ``SoulMan''.

\end{document}